\documentclass[10pt, conference, letterpaper]{IEEEtran}
\IEEEoverridecommandlockouts
\usepackage{cite}
\usepackage{amssymb,amsmath,amsthm,amsfonts}
\usepackage{algorithmic}
\usepackage[linesnumbered,ruled, boxed, lined, noend]{algorithm2e}
\usepackage{graphicx}
\usepackage{textcomp}
\usepackage{xcolor}
\usepackage[linesnumbered,ruled, boxed, lined]{algorithm2e}
\usepackage{url}

\def\BibTeX{{\rm B\kern-.05em{\sc i\kern-.025em b}\kern-.08em
    T\kern-.1667em\lower.7ex\hbox{E}\kern-.125emX}}

\allowdisplaybreaks

\DeclareMathOperator*{\argmin}{argmin}

\newcommand{\AoA}{\mathrm{AoA}}
\newcommand{\AoD}{\mathrm{AoD}}

\newcommand{\hiu}{\mathbf{h}_{i,u}}

\newcommand{\Huk}{\mathbf{H}_{u,k}}

\newcommand{\Cb}[1]{ \mathrm{C}_{#1} }

\begin{document}

\title{QTCAJOSA: Low-Complexity Joint Offloading and Subchannel Allocation for NTN-Enabled IoMT
}

\author{\IEEEauthorblockN{1\textsuperscript{st} Given Name Surname}
\IEEEauthorblockA{\textit{dept. name of organization (of Aff.)} \\
\textit{name of organization (of Aff.)}\\
City, Country \\
email address or ORCID}
}

\author{

\IEEEauthorblockN{ Alejandro Flores C.; Konstantinos Ntontin; Ashok Bandi; and Symeon Chatzinotas}
\IEEEauthorblockA{Interdisciplinary Centre for Security, Reliability and Trust (SnT), University of Luxembourg, Luxembourg.}
\IEEEauthorblockA{e-mails:\{alejandro.flores, kostantinos.ntontin, ashok.bandi, symeon.chatzinotas\} @uni.lu} \vspace{-2em}
}



\newcommand{\SAT}{\mathrm{SAT}}
\newcommand{\LEO}{\mathrm{LEO}}
\newcommand{\UAV}{\mathrm{UAV}}
\newcommand{\HAPS}{\mathrm{HAPS}}
\newcommand{\IoT}{\mathrm{IoT}}
\newcommand{\maxt}{\mathrm{max}}
\newcommand{\mint}{\mathrm{min}}
\newcommand{\Niot}{N_{\mathrm{IoT}}}

\newcommand{\Nuav}{N_{u}^{\mathrm{ULA}}}
\newcommand{\NuavU}{N_{u}^{\mathrm{UPA}}}
\newcommand{\Nhaps}{N_{h}}
\newcommand{\Nleo}{N_{s}}

\everydisplay{\small}

\maketitle

\begin{abstract}

In this work, we consider the resource allocation problem for task offloading from Internet of Medical Things (IoMT) devices, to a non-terrestrial network. The architecture considers clusters of IoMT devices that offload their tasks to a dedicated unmanned aerial vehicle (UAV) serving as a multi-access edge computing (MEC) server, which can compute the task or further offload it to an available high-altitude platform station (HAPS) or to a low-earth orbit (LEO) satellite for remote computing. We formulate a problem that has as objective the minimization of the weighted sum delay of the tasks. Given the non-convex nature of the problem, and acknowledging that the complexity of the optimization algorithms impact their performance, we derive a low-complexity joint subchannel allocation and offloading decision algorithm with dynamic computing resource initialization, developed as a greedy heuristic based on convex optimization criteria. Simulations show the gain obtained by including the different non-terrestrial nodes against architectures without them.

\end{abstract}


\section{Introduction}

As an enabler for the Heatlhcare 4.0 paradigm, the Internet of medical things (IoMT) allows for the acquisition and processing of biosignals recovered from patients, and shared with healthcare actors, such as hospitals and physicians. In the context of remote patient monitoring (RPM), some examples of IoMT devices include continuous glucose monitoring sensors~\cite{art:IoMT_CGM} and portable electrocardiograms~\cite{art:IoMT_ECG}. The continuous monitoring of biosignals of the patient allows for a more personalized healthcare by processing this data for detection or prediction of adverse patient conditions~\cite{art:IoMT_fever}.

IoMT devices are usually resource- and energy-constrained, thus processing their generated tasks at remote multiaccess edge computing (MEC) servers allows for their timely execution. Nevertheless, less than 45\% of the landmass of the Earth \cite{web:GSMA2024} and only 15\% of its surface has mobile coverage \cite{web:wef2024}, leaving many areas where IoMT devices cannot offload their tasks to the network. A cost-efficient solution to this problem is to make use of non-terrestrial networks (NTNs), which may comprise of unmanned aerial vehicles (UAVs), high altitude platform stations (HAPS) and satellites. 
Low-earth orbit (LEO) satellites are located 200-2000 km above Earth surface and have a beam footprint that can span up to a thousand kilometers, potentially having a large number of devices within service range.
LEO satellites are also characterized by limited computing resources, high mobility, and non-negligible propagation delays, making the coordination of a large number of devices challenging. A HAPS may offer a larger computing resource pool, while having a more limited coverage area. Moreover, UAVs can be deployed in an ad-hoc manner to gather IoMT-generated computing tasks, and computing them, or further offloading them for computation at a HAPS or a LEO.

The use of NTNs in task offloading has risen research interest in recent years. The problem of resource allocation in single-layer NTNs has been studied with UAVs~\cite{art:UAVOff_Zhang}, \cite{art:UAVDT_Hevesli}, HAPS~\cite{art:HAPSCaching_Ren}, or LEO satellites~\cite{art:RW_Ding1_2022},~\cite{art:RW_Lyu_2023}, as well as with multi-layer NTNs~\cite{art:RW_Waqar_2022}, 	\cite{art:RW_Chen_2023}. The solutions in the existing literature do not consider the algorithm runtime delay into the optimization problems, which further constraints the maximum delay of the tasks. Moreover, the solutions to the offloading problem usually consider a fixed initialization of variables in iterative solutions, which has the potential to bias the offloading decisions, or to not properly advertise the resources available at remote MEC servers.

Given the large distances across air and space nodes, it is important for delay-sensitive computing tasks, to design efficient algorithms to solve the problem of resource allocation in such networks. With this in mind, and different from related literature, we present a novel, low-complexity, joint subchannel allocation and offloading decision algorithm, with dynamic resource initialization to better advertise the resources of remote nodes to the tasks generated by IoMT devices. We propose this over a 3-layer NTN comprised of UAVs, a resource-rich HAPS and a large-coverage LEO satellite. Adding to the methodology of the literature, we consider the algorithm runtime delay in the total delay of the tasks. 

The remainder of this paper is structured as follows: Section~\ref{sec:SysModel} introduces the system model with the parameters of interest, whereas Section~\ref{sec:Problem} presents the optimization problem to be solved. Section~\ref{sec:Algorithm} develops the resource allocation algorithm. In Section~\ref{sec:Results} simulation results are presented, and finally, in Section~\ref{sec:Conclusions} the conclusions are drawn.

\section{System model}\label{sec:SysModel}

\begin{figure}[ht]
\vspace{-0.5em}
    \centering
    \includegraphics[width=0.85\linewidth]{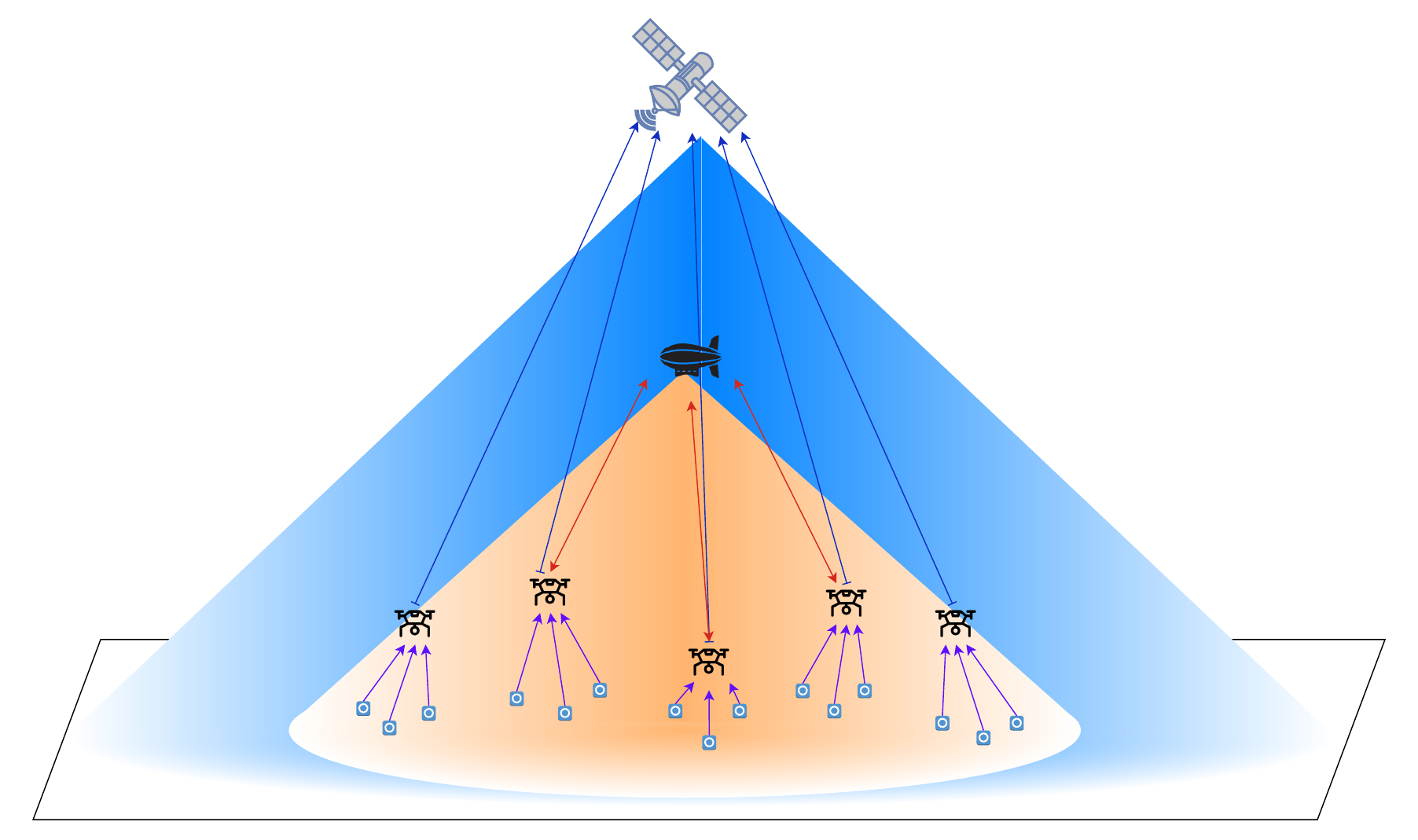}
    \caption{System model }
    \vspace{-0.7em}
    \label{fig:sysmodel}
\end{figure}

Consider a set of ground IoMT devices $\mathcal{I} = \{1,...,I\}$ separated in geographically distant clusters. A UAV, from set of UAVs $\mathcal{U} = \{1,...,U\}$ is assigned to each cluster, where the set of devices associated with UAV $u$ is $\mathcal{I}_u\subseteq\mathcal{I}$ with $|\mathcal{I}_u| = I_u$ and $\sum_{u\in\mathcal{U}}I_u = I$. In the system, there are a HAPS $h$ and a LEO satellite $s$ that act as MEC servers, with the HAPS acting as a coordinator across UAVs. The antenna elements at the HAPS and LEO satellite exhibit a smooth radiation pattern, such as in~\cite{art:RW_Ding1_2022}, which affects the range of the scanning beams of the arrays, determining their coverage areas. The UAVs are assumed to receive a minimum level of service, which allows them to exchange signaling information with both the HAPS and LEO satellite through channels in the same band. Each node is located in Cartesian coordinates $\mathbf{r}_k$$=$$[x_k, y_k, z_k]^T$, with $k\in \mathcal{I}\cup\mathcal{U}\cup\{h,s\}$. A diagram of the system is shown in Fig.~\ref{fig:sysmodel}.

In a snapshot of the system, the IoMT devices generate tasks irregularly to be processed. Each task is represented as $\pmb{\psi}_i = [ d_{i}, c_{i}, \tau_{i}^{\maxt} ]$, where $d_{i}$ is the number of bits in the task, $c_{i}$ is its computational density (in CPU cycles per bit), and $\tau_{i}^{\maxt}$ is its maximum tolerable delay (in seconds). Every IoMT device offloads its task to their local UAV, which acts as a MEC server and local coordinator for the cluster. The UAV can compute the task, or further offload it to the HAPS or to the LEO satellite for remote computation. The set of computing nodes is $\mathcal{R} = \mathcal{U}\cup\{h,s\}$.

A UAV communicates with its associated IoMT devices through orthogonal subchannels $\mathcal{B} =\{ 1,...,B  \}$, 
of bandwidth $B_u$. 
In remote areas the spectrum is not used extensively, for which there is no need for high frequency reuse, allowing for orthogonal access. Thus,
each IoMT device occupies one orthogonal subchannel, having no inter-IoMT device interference. Due to the directivity of the antenna arrays and the geographical separation between clusters, we assume no interference between clusters. For purposes of communication and signaling, the IoMT devices are single-antenna devices, while UAVs have a uniform linear arrays (ULAs) of $\Nuav$ antenna elements. For communication between UAVs and the HAPS or LEO satellite, the UAVs have an upward-facing uniform planar array (UPA) of $\NuavU$ antenna elements, while the HAPS and the LEO satellite have downward facing UPAs of $\Nhaps$ and $\Nleo$ antenna elements respectively. They communicate through orthogonal subchannels, having no interference.

\subsection{Communication model}
\subsubsection{IoMT - UAV}

We consider that a given UAV applies a maximum-ratio combiner to the signal from IoMT device $i$ communicating task $\pmb{\psi_i}$. 
Assuming a quasi-static channel in the S band, the instantaneous channel capacity of the link between IoMT device $i$ to UAV $u$ through subchannel $b$ is given as 
{\small
\begin{align}\label{eq:Riu}
    R_{i,u}^b =  B_u \log_2\left( 1 + \frac{||\hiu^{b}||^2p_{i,u}^{b}}{B_uN_0} \right)
\end{align}
}where $p_{i,u}^{b}$ is the transmit power, $N_0$ is the noise power per unit bandwidth, and $\mathbf{h}_{i,u}^{b}$$=$$\sqrt{L_{i,u}^{-1}}\Bar{\mathbf{h}}^{b}_{i,u}$ is the air-to-ground (A2G) channel response between $i$ and $u$ through subchannel $b\in\mathcal{B}$, where $L_{i,u}$ is the pathloss between $i$ and $u$ given as in~\cite{art:Hourani_UAVChannels}, and the small-scale fading component $\Bar{\mathbf{h}}^{b}_{i,u}$ is modeled as Rician fading as in~\cite{art:Cheng_RicianChans}, for which the expression for the corresponding Rician factor can be found in~\cite{art:Kfactor_Azari_2018}. Let $\rho_{i,u}^{b}$ be a binary variable indicating if subchannel $b$ is allocated to device $i$ by UAV $u$ ($\rho_{i,u}^{b}=1$), or not ($\rho_{i,u}^{b}=0$). Then, the instantaneous channel capacity can be written as $R_{i,u} = \sum\limits_{b\in\mathcal{B}} \rho_{i,u}^{b} R_{i,u}^b $.

\subsubsection{UAV - HAPS/LEO}
Having more simplified coordination between UAVs and the HAPS/LEO satellite, we assume that the corresponding links in the Ka band are orthogonal of bandwidth $B_h$ and $B_s$, with channel matrices $\Huk\in\mathbb{C}^{N_k \times \NuavU}$ with $k\in\{h,s\}$. Given the spatial sparsity of the links between the upward facing UPAs of the UAVs, and the downward facing UPA of the HAPS/LEO satellite, these channels are modeled as LoS channels undergoing free-space pathloss, as 
\begin{align}
    \Huk = \left( \frac{\lambda_k g_k(\theta_{u,k})\sqrt{G_{u,k}}}{4\pi ||\mathbf{r}_u - \mathbf{r}_k||_2} \right) \mathbf{a}^{\AoA}_{k,u}(\mathbf{a}^{\AoD}_{u,k})^H,
\end{align}
where $\mathbf{a}^{\AoA}_{k,u}$ and $\mathbf{a}^{\AoD}_{u,k}$ 
are beamsteering vectors of the corresponding angle of arrival and angle of departure, $G_{u,k}$ includes the atmospheric losses due to weather phenomena in the troposphere, and $g_k(\theta_{u,k})$ models the normalized radiation pattern of the antenna elements of node $k$. For the LEO, we consider horn-type antenna elements as commercially available \cite{web:LEOantenna}, and we choose a Bessel-based radiation pattern as in~\cite{art:AntPat_Scwarz_2019}. For the HAPS, we consider a patch antenna array as considered in the literature \cite{art:HAPSPatch1_Hee}, and in commercial models \cite{web:HAPSantenna} and consider a cosine radiation pattern.

The LoS-MIMO channel is rank-deficient, having only array gain, for which a singular-value decomposition 
is performed to obtain the precoder employed by the UAV and the combiner employed by the HAPS or LEO satellite. Then, the capacity of the channel is 
\begin{align}\label{eq:Rih}
    R_{u,k} &= B_k \log_2\left( 1 + p_{u,k}\frac{\sigma_{k}^2}{B_kN_0} \right)
\end{align}
where $\sigma_{k}$ is the non-zero singular value of the channel matrix. 
Let $\beta_{u,k}^i$ be a binary variable indicating if task $i$ is offloaded from UAV $u$ to the node $k$ ($\beta_{u,k}^i=1$) or not ($\beta_{u,k}^i=0$).

\subsection{Delay model}
We consider the delay by the transmission and computing of the task, and the delay by the signaling and algorithm runtime. The delay of reporting the results is typically not considered due to their small size compared to the size of the tasks.

\subsubsection{Transmission delay} This delay is induced by the transmission of the tasks between nodes.

\paragraph{IoMT - UAV}
The transmission delay experienced by offloading task $\pmb{\psi}_i$ from device $i$ to UAV $u$ is given as
\begin{align}
    \tau_{i,u} = \frac{d_i}{ R_{i,u}} = \frac{d_i}{\sum\limits_{b\in\mathcal{B}} \rho_{i,u}^{b} R_{i,u}^b}.
\end{align}

\paragraph{UAV - HAPS/LEO}
The transmission delay of offloading the required tasks from UAV $u$ to node $k\in\{h,s\}$, is
\begin{align}
    \tau_{u,k} &= \frac{ \sum\limits_{i\in\mathcal{I}_u} \beta_{u,h}^i d_i}{R_{u,h}}+ 2\frac{||\mathbf{r}_{k} - \mathbf{r}_{u}||_2}{c}.
\end{align}
The second term is the round-trip propagation delay, which is negligible for the UAV-HAPS link, but it is considered for the UAV-LEO link due to the large distances considered.

\subsubsection{Computing delay} 
If node $k\in\mathcal{R}$ computes task $\pmb{\psi}_i$, the corresponding computing delay is given as 
\begin{align}
    \tau_i^k = \frac{  d_{i} c_{i} }{f_i^k} 
\end{align}
where $f_i^k$ are the resources allocated for task $\pmb{\psi}_i$ by node $k$.

\subsubsection{Algorithm Runtime Delay}
For real-time task offloading, the runtime of the resource allocation algorithm must be considered. To compute this, the number of basic operations in the algorithm (additions, multiplications and divisions\footnote{Usually, divisions are considered with a different number of cycles, but for simplicity we will include them with the basic operations.}) is $o_{op}$, while the number of CPU cycles per operation is $c_{op}$, which depends on parameters such as the type of processor, number of cores and operation units~\cite{the:Processors_Lovelly}. Then, the algorithm runtime delay for centralized execution at node $k$ is given as 
\begin{align}\label{eq:runtime_delay}
    \tau_{op} &= \frac{o_{op} c_{op}}{F_k^{\maxt}} + \tau_{k}^{\mathrm{RTT}}.
\end{align}
$F_k^{\maxt}$ is the total computing resources of node $k$ and $\tau_{k}^{\mathrm{RTT}}$ is the maximum round trip time between ground nodes and $k$. The transmission times are not considered since the signaling overhead is assumed to be of much lower size.

\subsection{Weighted Sum Delay}\label{sec:Perf_Met}
 
The delay minimization may prioritize tasks with longer delay constraints, so to introduce fairness across tasks, a weighted sum delay 
is considered as 
{\small
\begin{align}
    \nonumber \Hat{\tau} &= \sum\limits_{i\in\mathcal{I}} \frac{ 1}{\tau_{i}^{\maxt}} \left( \tau_{op} + \tau_{i,u} + \sum\limits_{k\in\{h,s\}}\beta_{u,k}^{i} \tau_{u,k}  \right. \\
     &\left.  + \left(1 - \beta_{u,h}^{i} - \beta_{u,s}^{i}\right)\tau_i^u + \sum\limits_{k\in\{h,s\}}\beta_{u,k}^{i} \tau_i^k \right)
\end{align}
}

We can write $\Hat{\tau} = \sum\limits_{i\in\mathcal{I}} \frac{ 1}{\tau_{i}^{\maxt}} \tau_{i} $ with $\tau_i = \tau_{op}  + \Bar{\tau}_i^{u} + \beta_{u,h}^{i}\Bar{\tau}_i^{h} + \beta_{u,s}^{i}\Bar{\tau}_i^{s}$, where each term represents the delay of offloading and computing a task at a certain node. 
Here $\Bar{\tau}_i^{u} = \tau_{i,u} + \tau_i^{u}$, and $\Bar{\tau}_i^{k} = \tau_{i,u} + \tau_{u,k} + \tau_i^{k} - \tau_i^{u}$ for $k\in\{h,s\}$.

\section{Problem formulation}\label{sec:Problem}

Let  $\pmb{P}_u = [\pmb{\rho}_1,...,\pmb{\rho}_I]$ with $\pmb{\rho}_i = [\rho_{i,u}^{1}, ..., \rho_{i,u}^{B}]^{T}$, $\pmb{B}_{u} = [\pmb{\beta}_{u,h},\pmb{\beta}_{u,s}]$, with $\pmb{\beta}_{u,k} = [\beta_{u,k}^{1},...,\beta_{u,k}^{I_u}]^{T}$ for $k\in\{h,s\}$, and $\mathbf{F} = [\mathbf{f}^{1},...,\mathbf{f}^{U}, \mathbf{f}^{h},\mathbf{f}^{s}]$, with $\mathbf{f}^{k} = [f_1^k,...,f_I^k]^{T}$ for $k\in\{u,h,s\}$. Defining  $\mathcal{S} = \{  \{ \pmb{P}_u \}_{u\in\mathcal{U}},  \{\pmb{B}_{u}  \}_{u\in\mathcal{U}}, \mathbf{F}  \} $, the optimization problem to be solved is formulated as follows:
\begin{subequations}\label{eq:mainOpt_all1}
{
\small
\begin{alignat}{3}\label{eq:mainOpt_1}
\mathcal{P}:\;\;\; &\min_{\mathcal{S}} & &\Hat{\tau}  \\ 
\label{eq:mainOpt_7}        &\text{s.t.}   &\Cb{1}:\quad&  \tau_i \leq \tau_{i}^{\maxt},  \;\;  \forall i\in\mathcal{I}\\
\label{eq:mainOpt_6}        &   &\Cb{2}:\quad&  \sum\limits_{i\in\mathcal{I}} f_i^k \leq F_k^{\maxt},  \;\;  \forall k\in\mathcal{R}\\
\label{eq:dyn_off_loc_2}    &   &\Cb{3}:\quad&     \sum\limits_{b\in\mathcal{B}} \rho_{i,u}^{b} \leq 1,  \;\; \forall u\in\mathcal{U}, \forall i\in\mathcal{I}_u \\
\label{eq:dyn_off_loc_3}    &   &\Cb{4}:\quad&    \sum\limits_{i\in\mathcal{I}} \rho_{i,u}^{b} \leq 1,  \;\; \forall u\in\mathcal{U}, \forall b\in\mathcal{B}         \\
\label{eq:dyn_off_uav-hs}   &   &\Cb{5}:\quad&    \beta_{u,s}^{i} + \beta_{u,h}^{i} \leq \sum\limits_{b\in\mathcal{B}} \rho_{i,u}^{b},  \;\; \forall u\in\mathcal{U}, \forall i\in\mathcal{I}_u   \\
\label{eq:mainOpt_9}        &   &\Cb{6}:\quad&  \beta_{u,h}^{i}, \beta_{u,s}^{i} \in \{0,1\},  \;\;  \forall u\in\mathcal{U}, \forall i\in\mathcal{I} \\
\label{eq:mainOpt_13}       &   &\Cb{7}:\quad&  \rho_{i,u}^{b} \in \{0,1\}, \;\;  \forall u\in\mathcal{U}, \forall i\in\mathcal{I}_u, \forall b\in\mathcal{B}         \\
\label{eq:posComp}          &   &\Cb{8}:\quad&    f_i^k \geq 0,  \;\;   \forall i\in\mathcal{I},  k\in\mathcal{U}\cup\{h,s\} 
\end{alignat}
}\end{subequations}
where $\Cb{1}$ constraints the maximum delay of every task, $\Cb{2}$ constraints the maximum computing resources at the remote nodes, $\Cb{3}$ constraints IoMT devices to access their local UAVs through at most one subchannel, $\Cb{4}$ constraints the IoMT-UAV subchannels to be occupied by at most one IoMT device, $\Cb{5}$ indicates a UAV can offload a task to the HAPS/LEO satellite only if the task has been offloaded to it, and $\Cb{6}$ and $\Cb{7}$ constraints the decision variables to be binary.

\section{Proposed Resource Allocation Algorithm}\label{sec:Algorithm}

To solve $\mathcal{P}$ we propose a two-stage block-coordinate descent algorithm that performs the offloading decision of every task, followed by the optimal resource allocation at every computing node. The offloading decision is designed with dynamic resource initialization, based on optimality conditions. For this reason, the computing resource optimization is presented first.

\subsubsection{Remote Computing Resources}\label{sec:Rem_F}

We can express $\mathcal{P}$ in terms of the remote computing resources, keeping the offloading variables fixed. This problem can be decoupled per remote node, and the maximum delay constraint can be transformed into a minimum computing resources constraint. Defining $\mathcal{I}^k$ as the set of tasks computed in remote node $k$, each subproblem can be written as
\begin{subequations}
{\small
\begin{alignat}{3}
\label{eq:optprob_f_1_orig}\mathcal{P}_{k}^{\text{R-F}}:\;\;\; &\min_{\{f_i^k\}_{i\in\mathcal{I}^k}} & & \sum\limits_{i\in\mathcal{I}^k}\frac{d_{i} c_{i}}{\tau_{i}^{\maxt}} \frac{ 1  }{f_i^k}    & \\ 
\label{eq:optprob_f_2_orig}&\text{s.t.}   &\quad&  \sum\limits_{i\in\mathcal{I}^k} f_i^k \leq F_{k}^{\maxt}          & \\
\label{eq:optprob_f_3_orig}&   &\quad&  f_i^k  \geq \frac{d_{i} c_{i}}{\tau_{i}^{\maxt} - \tau_{i\rightarrow k}} = f_{i}^{k,\text{min}}      \qquad \forall i\in\mathcal{I}^k    & \\
\label{eq:optprob_f_4_orig}&   &\quad&  f_i^k  \geq 0     \qquad \forall i\in\mathcal{I}^k     & 
\end{alignat}
}\end{subequations}
where $\tau_{i\rightarrow k}$ is the delay of task $\pmb{\psi}_i$ to arrive to node $k$, given as $\tau_{i\rightarrow u} = \tau_{i,u}$ and $\tau_{i\rightarrow k} = \tau_{i,u} + \tau_{u,k} $ for $k\in\{h,s\}$, where $\mathcal{I}_u^k$ are the set of tasks offloaded from UAV $u$ to $k\in\{h,s\}$.

The solution to problem $\mathcal{P}_{k}^{\text{R-F}}$ without considering the minimum-resources constraint, is given in closed-form as 
\begin{align}\label{eq:optf_remote}
    f_i^{k,*} &= \frac{\sqrt{ \frac{d_{i} c_{i}}{\tau_{i}^{\maxt}} }}{\sum\limits_{i\in\mathcal{I}^k}  \sqrt{ \frac{d_{i} c_{i}}{\tau_{i}^{\maxt}} }} F_{k}^{\maxt} \qquad \forall i\in\mathcal{I}^k
\end{align}
The details are omitted due to page constraints. If a task is not feasible at the optimal point, the algorithm will allocate their minimum resources, if possible. If not, the task is computed at the local UAV. If the task can not be feasibly computed at the UAV, it allocates $f^{u,\mathrm{min}}_i$ to every task in ascending order, until it is not possible for a given task. The remaining resources are given to the remaining tasks equally in a best-effort solution.

The remote computation resource allocation sets computation resources for tasks not allocated to the remote node to zero, which stops a task from changing their computing node. To avoid this, the remote computing resource allocation problem is solved after the offloading decision problem.

\subsubsection{Quadratic Transform-enabled Computing-Aware Joint Offloading and Subchannel Allocation (QTCAJOSA)}\label{sec:QTCAJOSA}

We write $\mathcal{P}$ in terms of the subchannel allocation and offloading decision variables, while keeping the other variables fixed, as
\begin{subequations}\label{eq:GBOpt_all2}
{\small
\begin{alignat}{3}\nonumber
\mathcal{P}^{\text{B-D}}: &\min_{ \substack{\{ \pmb{B}_u \}_{u\in\mathcal{U}} \\ \{\pmb{P}_{u}  \}_{u\in\mathcal{U}} } } & &\sum_{i\in\mathcal{I}}\left(\sum_{b\in\mathcal{B}}\rho_{i,u}^{b} \frac{\upsilon_{u,b}^{i}}{\tau_{i}^{\maxt}} \right. & \\ 
\label{eq:GBOpt2_1}&  & &+ \left.\sum\limits_{k\in\{h,s\}}\beta_{u,k}^{i} \frac{1}{\tau_{i}^{\maxt}}\left( \frac{\sum\limits_{j\in\mathcal{I}}  \beta_{u,k}^{j} d_j}{R_{u,k}}  +  \upsilon_{k}^{i}\right)\right) & \\
\label{eq:GBOpt21_2} &\text{s.t.}   &\quad&  \Cb{1}, \Cb{3}-\Cb{7} & 
\end{alignat}
}\end{subequations}
where $\upsilon_{u,b}^{i} = \tau_i^u + \frac{d_i}{R_{i,u}^{b}}$, $\upsilon_{h}^{i} =  \tau_i^h - \tau_i^u$ and $\upsilon_{s}^{i} =  \tau_i^s - \tau_i^u + \frac{2||\mathbf{r}_{s} - \mathbf{r}_{u}||_2}{c}$. To decouple the objective function, we apply the following algebraic transformation to the quadratic cross-term
{\small
\begin{align}
    \frac{\beta_{u,k}^{i}\sum\limits_{j\neq i}d_j\beta_{u,k}^{j}}{R_{u,k}} &= \frac{\left(\sum\limits_{j\neq i}d_j\beta_{u,k}^{j} + \beta_{u,k}^{i}\right)^2 - \left(\sum\limits_{j\neq i}d_j\beta_{u,k}^{j} - \beta_{u,k}^{i}\right)^2}{4R_{u,k}}
\end{align}}
Then, we apply the quadratic transform (QT)~\cite{art:QT_Shen_2018} to each of the terms as
{\small
\begin{align}
    \frac{\left(\sum\limits_{j\neq i}d_j\beta_{u,k}^{j} \pm \beta_{u,k}^{i}\right)^2}{R_{u,k}} &\Rightarrow 2z_{\pm,i} \left| \sum\limits_{j\neq i}d_j\beta_{u,k}^{j} \pm \beta_{u,k}^{i} \right| - z_{\pm,i}^{2} R_{u,k} 
\end{align}}
where $z_{\pm,i} = \frac{1}{R_{u,k}^{(n)}}\left| \sum\limits_{j\neq i}d_j\beta_{u,k}^{j,(n)} \pm \beta_{u,k}^{i,(n)} \right|$ is computed from the results of the previous iteration. Assuming that previously allocated tasks cannot be removed from their allocated nodes, we can remove the absolute value, and it becomes
{\small
\begin{align}
    \nonumber\frac{\beta_{u,k}^{i}\sum\limits_{j\neq i}d_j\beta_{u,k}^{j}}{R_{u,k}} &=  \left(\frac{ \beta_{u,k}^{i,(n)}}{R_{u,k}^{(n)}} \right)\sum\limits_{j\neq i}d_j\beta_{u,k}^{j} + \left(\frac{ \sum\limits_{j\neq i}d_j\beta_{u,k}^{j,(n)}}{R_{u,k}^{(n)}}\right)\beta_{u,k}^{i} \\
    \label{eq:QT_beta}&- \left(\frac{\beta_{u,k}^{i,(n)}\sum\limits_{j\neq i}d_j\beta_{u,k}^{j,(n)}  }{\left(R_{u,k}^{(n)}\right)^{2}}\right) R_{u,k} 
\end{align}}

Thus, the objective function is decoupled per-task such that $\beta_{u,k}^{i}$ is weighted by the cost on the task $i$ and on the impact on all of the already allocated tasks. The optimization problem over the offloading decision variables per UAV $u$ is given as

{
\footnotesize
\begin{subequations}
{
\begin{alignat}{3}
\mathcal{P}^{\text{B-D}}_u:\;\;\; &\min_{ \pmb{B}_u, \pmb{P}_u } & & \sum\limits_{i\in\mathcal{I}_{u}} \left(\sum\limits_{b\in\mathcal{B}} \rho_{i,u}^{b} \upsilon_{u,b}^{i,(n)} + \beta_{u,h}^{i}\upsilon_{h}^{i,(n)} +  \beta_{u,s}^{i}\upsilon_{s}^{i,(n)}\right)   & \\ 
&\text{s.t.}   &\quad&  \Hat{\Cb{1}}, \Hat{\Cb{3}}-\Hat{\Cb{7}} & 
\end{alignat}
}\end{subequations}}
where $\Hat{\Cb{}}$ is constraint $\Cb{}$ only for the nodes $i\in\mathcal{I}_u$, and
{\small
\begin{align}\label{eq:ups_k}
    \upsilon_{k}^{i,(n)} &=  \frac{1}{\tau_{i}^{\maxt}}\left(\upsilon_{k}^{i} + \frac{d_i}{R_{u,k}} + \frac{ \sum\limits_{j\neq i}d_j\beta_{u,k}^{j,(n)}}{R_{u,k}}\right) +   \frac{ d_i\sum\limits_{j\neq i}\frac{\beta_{u,k}^{j,(n)}}{\tau_{j}^{\maxt}}}{R_{u,k}} 
\end{align}}
for $k\in\{h,s\}$. With the problem expressed in this form, we define a greedy algorithm that allocates tasks one by one, checking the best allocation at each step. The solution to this problem depends on the initializations of the $f_{i}^{k}$ variables. If $f_{i}^{k} < f_i^u$ for a given $i$, the task will not be offloaded. This happens even if the total pool of resources that node $k$ can provide to task $i$ is much larger. For a large number of tasks in the system, this can be true for every task, and no offloading takes place. To address this issue, we propose two measures:
\paragraph{Non-orthogonal Computing Resource Initializations}
Let $F_{k,u}^{\maxt}$ be the total resources available at remote node $k$ allocated to UAV $u$. We consider, $F_{k,u}^{\maxt} = F_{k}^{\maxt}$, thus $f_{i}^{k}$ is initialized considering that the serving UAV of $i$ can make use of all the resources of node $k$. This results in overlapping resources across UAVs, thus after each allocation, the best task across the ones from each UAV is the one allocated.

\paragraph{Dynamic Computing Resource Initialization}
After each allocation, the computing resources at each remote node $k\in\mathcal{R}$ are initialized only for the non-allocated tasks.
Let $\Hat{\mathcal{I}}\subseteq\mathcal{I}$ be the set of tasks not yet allocated. At the start of each round the computation resources $f_{i}^{k}$ for $i\in\Hat{\mathcal{I}}_u$ is initialized as \eqref{eq:optf_remote}, for $k\in\mathcal{R}$. This is implemented for the UAV and LEO satellite and not the HAPS, given that the HAPS has a larger pool of resources that could deplete early with this method.

Then, considering checks for maximum delay feasibility, we present the quadratic transform-enabled computing-aware joint offloading and subchannel allocation (QTCAJOSA) algorithm in Algorithm~\ref{alg:3c_QTCAJOSA_HAPS}. 
The total number of operations for QTCAJOSA is $I( 6(I^2) + 6 I B + 100I + 3B + 100 )$, and for the remote computing resources allocation algorithm is $(U+2)(2(I_u^2) + 24 I_u + 15)$\footnote{We assume that additions, multiplications and divisions count as elementary operations and that the number of elementary operations in other familiar functions is 10, to not consider a specific methodology or CPU architecture.}.

\SetKwBlock{RepeatE}{repeat}{}
\begin{algorithm}[t]
{\small
    \caption{ QTCAJOSA }\label{alg:3c_QTCAJOSA_HAPS}
    \ForEach{$u\in\mathcal{U}$}{
        Initialize $\pmb{\beta}_u^h = \mathbf{0}_{I_u\times 1}$, $\pmb{\beta}_u^s = \mathbf{0}_{I_u\times 1}$ and $\pmb{\rho}_u = \mathbf{0}_{I_u\times B_u}$\;
        Define $\mathbf{v}_u^h = \mathbf{v}_u^s = \mathbf{v}_u^u = \mathbf{1}_{I_u\times 1}$ and $\mathbf{V}_u = \mathbf{1}_{I_u\times B_u}$\;
         Initialize $\Hat{\mathcal{I}}_u = \mathcal{I}_u$\;
     }
    \RepeatE{
        \ForEach{$u\in\mathcal{U}$}{
        Initialize $f_{i}^u$ as in \eqref{eq:optf_remote} $\forall i\in\Hat{\mathcal{I}}_u$. \;
        Compute $\mathbf{U}_u$ where $\mathbf{U}_u(i,b) = \upsilon_{u,b}^{i}$\;
        Compute $\Bar{\mathbf{U}}_u^u := \mathbf{U}_u \odot \mathbf{V}_u \odot \mathbf{v}_u^u $ (column-wise)\; 
        \ForEach{$k\in\{h,s\}$}{
            Initialize $f_{i}^k$ as in \eqref{eq:optf_remote} $\forall i\in\Hat{\mathcal{I}}_u$. \;
            Compute $\Bar{\mathbf{u}}_u^k(i) = [\upsilon_{k}^{i},...,\upsilon_{k}^{I_u}]^T$\;
            Compute $\Bar{\mathbf{u}}_u^k := \Bar{\mathbf{u}}_u^k \odot \mathbf{v}_u^k$\;
            Compute $\Bar{\mathbf{U}}_u^k = \mathbf{U}_u + \Bar{\mathbf{u}}_u^k$ (column-wise)\;
        }
        }
        \textbf{if}  $\min\{ \Bar{\mathbf{U}}_{u}^{k}\}_{k\in\mathcal{U}\cup\{h,s\}} == +\infty$, \textbf{then} \Return{}\;
        Choose $(i^*,u^*,b^*, k^*) = \argmin\{ \Bar{\mathbf{U}}_{u}^{k}(i,b)\}_{k\in\mathcal{U}\cup\{h,s\}}$ \;
        \If{$\tau_{i^*} > \tau_{i^*}^{\maxt}$}{
            Set $f_{i^*}^{k^*} = F_{u^*,k^*}^{\maxt}$\;
            \If{$\tau_{i^*} > \tau_{i^*}^{\maxt}$}{
                Set $\mathbf{v}_{u^*}^{k^*}(i^*) = \infty$\;
                Go back to 2 \;
            }
        }

        Set $\mathbf{V}_{u^*}(i^*, :)$ and $\mathbf{V}_{u^*}(:, b^*)$ to $+\infty$\;
         Set $\mathbf{v}_{u^*}^{k^*}(i^*)$ to $+\infty$\;
         Set $\Hat{\mathcal{I}}_{u^*} := \Hat{\mathcal{I}}_{u^*} \setminus \{ i^* \}$\;
         Set $\rho_{i^*,u^*}^{b^*} = 1$ \;
         \textbf{if} $k^*\in\{h,s\}$, \textbf{then} Set $\beta_{u^*,k^*}^{i^*}=1$\;
         Update $F_{u^*}^{\maxt}:= F_{u^*}^{\maxt} - f_{i^*}^{k^*}$\;
    }{}
    }
\end{algorithm}

\section{Results}\label{sec:Results} 

To evaluate the proposed algorithm and architecture, we develop simulations comparing our results with benchmarks where there is no LEO satellite present (No LEO), no HAPS present (No HAPS), and where the offloading decision problem is not adaptive (Non-adaptive), having the computing resources initialized in a non-overlapping manner at the start and do not change throughout. 
Following 3GPP recommendations~\cite{rep:3gpp_38.101-5}, we consider the carrier frequency for the UAV-LEO/HAPS link in the Ka band, as $28$GHz with bandwidths $B_s=200$MHz and $B_h=100$MHz. For the IoMT-UAV links we assume LTE-M enabled IoMT devices working in LTE channel~\cite{rep:3gpp_21.914}, thus we choose $B_u = 1.4$MHz and $B=14$ subchannels over the carrier at $2.1$GHz. The altitude of the UAVs is 120m, according to the maximum altitude set by the European Union Aviation Safety Agency~\cite{law:EU_UAV}. The altitude of the LEO satellite and the HAPS are 500 km and 20 km, both at horizontal coordinates $[0,0]$ at the time of algorithm execution. We consider clear-sky conditions such that $G_{u,h} = G_{u,s} = 1$. For the tasks, we consider $\tau_i^{\maxt}=30$s, $c_i=100$ and $d_i$ to follow an uniform distribution $U(100\text{Kb}, 10\text{Mb})$, where larger sizes represent tasks gathered from IoMT devices such as wearable ultrasound devices for image processing, or continuous high-frequency time-series processing from wearable ECGs, and smaller sizes represent tasks such as time-series processing from IoMT devices such as wireless stethoscopes or wearable pulse oximeters.

\begin{figure}[!h]
    \centering
    \includegraphics[width=0.8\linewidth]{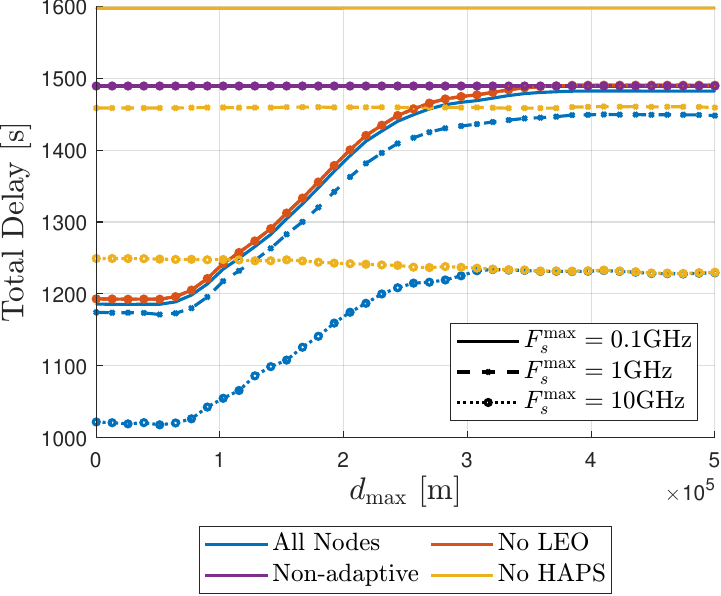}
    \vspace{-0.5em}
    \caption{Total delay across tasks for varying ground maximum distance.}\vspace{-1.5em}
    \label{fig:res_dmax}
\end{figure}
Fig.~\ref{fig:res_dmax} shows the total delay for varying maximum ground distance of an IoMT device to the center of coordinates $d_{\max}$, where curves are presented for different values of $F_s$.
The delay obtained by using the HAPS and the LEO satellite as MEC servers is smaller than the one obtained by not using either. There is a single curve for the no-LEO case, since only the HAPS is present and not affected by the resources at the LEO, as well as for the non-adaptive case because the initialized resources at the LEO are so small that it is never considered for offloading. The no-LEO curve is lower at smaller values of $d_{\max}$, since the nodes are within coverage of the HAPS, but also, it is closer to the all-nodes case when there are fewer resources at the LEO. As $d_{\max}$ increases, the UAVs fall outside coverage of the HAPS, and the transmission delay increases. As the resources available at the LEO satellite increase, the gap between the all-nodes case and the no-HAPS case decrease, since the resources at the LEO satellite are further utilized to compute the tasks not done so at the HAPS.

\begin{figure}[!h]
    \centering
    \includegraphics[width=0.8\linewidth]{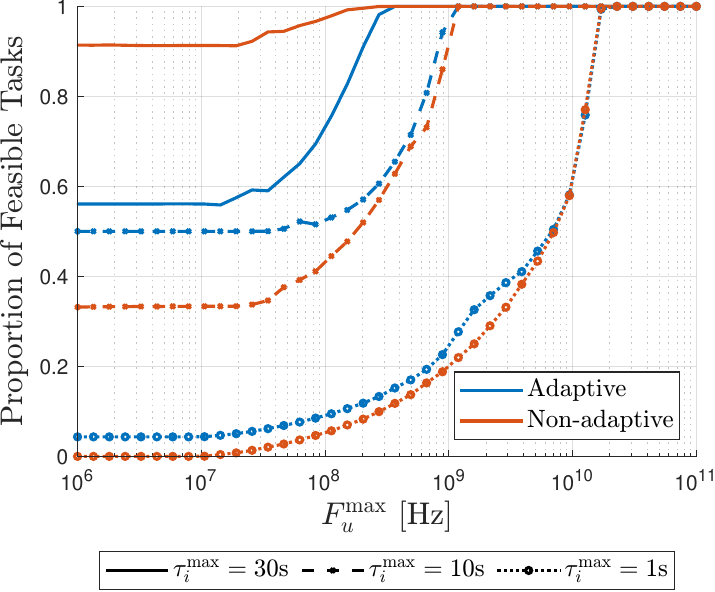}
    \vspace{-0.5em}
    \caption{Proportion of feasible tasks for varying UAV available resources.}\vspace{-0.5em}
    \label{fig:res_Fu}
\end{figure}

Fig.~\ref{fig:res_Fu} presents the proportion of feasible tasks in the system for varying $F_{u}^{\maxt}$, with curves for the proposed adaptive algorithm, and the non-adaptive version of the algorithm. The amount of feasible tasks in the system increases with the available resources at the UAVs, since they can compute more tasks. When the delay constraints are tighter, the proposed adaptive algorithm enables the fulfillment of more tasks feasibly due to its dynamic initialization and broadcasting of resources. If the delay constraints are looser, most of the tasks can be feasibly computed at the HAPS with the first initialization of resources, making the non-adaptive algorithm achieve higher number of feasible tasks than the adaptive algorithm, which may unevenly allocate resources across tasks.

\begin{figure}[!h]
    \centering
    \includegraphics[width=0.8\linewidth]{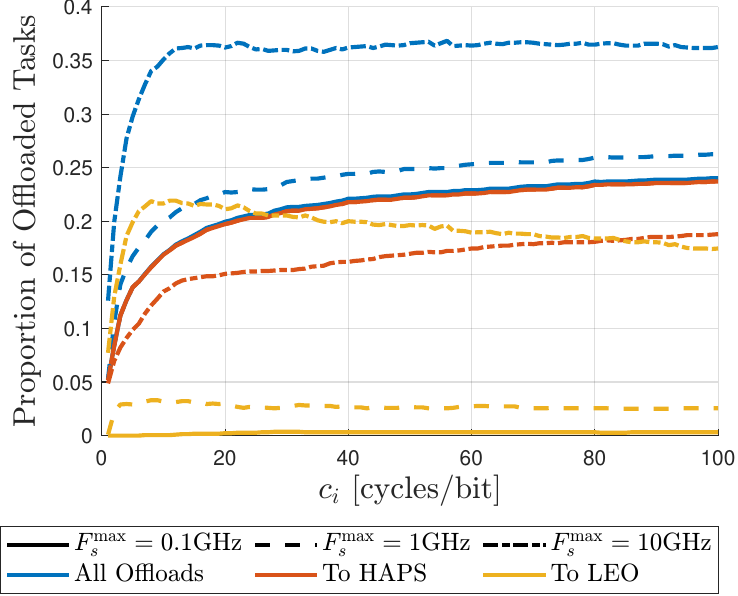}
    \vspace{-0.5em}
    \caption{Proportion of offloaded tasks per node for varying computation density.}\vspace{-1.5em}
    \label{fig:res_ci}
\end{figure}
Fig.~\ref{fig:res_ci} presents the proportion of tasks in the system offloaded to the HAPS or the LEO satellite for varying $c_i$, with three different values of $F_{s}^{\maxt}$. As the computing density increases, so does the offloading of the tasks to remote nodes because the computing density penalizes the computing delay at the UAV, but not the transmission delay of the task to the remote nodes. As the resources at the LEO satellite increase, more tasks are offloaded to it. Nevertheless, for the case of $F_s^{\maxt}=10$GHz, many more tasks are offloaded to the LEO rather than the HAPS, despite having the same amount of resources. This occurs because the HAPS initializes its resources at every step for all of the original tasks to avoid early depletion of resources, whereas for the LEO it offloads only to non-allocated tasks.

\section{Conclusions}\label{sec:Conclusions}

This work studied the problem of resource allocation for task offloading of IoMT-generated tasks to a NTN consisting of dedicated UAVs per cluster of ground devices, a common HAPS and a common LEO satellite acting as MEC servers. The solution was given as a low-complexity algorithm that optimizes the joint subchannel allocation and offloading decision, while having a dynamic computing resource initialization. It was shown how the HAPS offers more resources within its coverage, the LEO provides wide coverage with shared resources, and the UAV serves its specific IoMT devices. Adaptive computing resource initialization in offloading decisions reduces delay by better using smaller resource pools, while non-adaptive initialization may suit very large shared pools. Future work could explore optimizing other system resources and alternative algorithm distribution methods.

\section*{Acknowledgement}
This work received funding from Horizon Europe under the Marie Sklodowska Curie actions: ELIXIRION (GA 101120135).

\bibliographystyle{IEEEtran}

\bibliography{readme.bib}

\vspace{12pt}

\end{document}